\documentclass[twocolumn,twoside,slac_two]{revtex4}
\usepackage{graphicx}
\usepackage{fancyhdr}
\begin{document}

\title{GeV-TeV Blazar Population Studies}

%

\author{G.~D. \c{S}ent\"{u}rk}
\affiliation{Department of Physics, Columbia University, 550 West 120th Street, New York, NY 10027, USA}
\author{M. Errando, R. Mukherjee}
\affiliation{Department of Physics \& Astronomy, Barnard College, Columbia University, 3009 Broadway, New York, NY 10027, USA}

\author{M. B\"{o}ttcher, P. Roustazadeh}
\affiliation{Astrophysical Institute, Department of Physics and Astronomy, Ohio University, Athens, OH 45701, USA}

\author{P. Coppi}
\affiliation{Department of Astronomy, Yale University, P.O. Box 208101, New Haven, CT 06520, USA}

\begin{abstract}
The synergy between the \emph{Fermi}-LAT and ground-based Cherenkov telescope arrays gives us the opportunity for the first time to characterize the high energy emission (100 MeV -- 10 TeV) from more than 30 blazars. In this study we performed a \emph{Fermi}-LAT spectral analysis for all TeV-detected blazars and combined it with archival TeV spectra. Our results for low synchrotron-peaked BL Lacs (LBL) show hints of absorption features in the GeV band that could be interpreted as internal opacity at the source. We note that simple or broken power laws cannot describe all the observed GeV-TeV spectra and more complex spectral shapes seem required.

\end{abstract}

\maketitle

\thispagestyle{fancy}


\section{INTRODUCTION}
Active Galactic Nuclei (AGN) are extreme extragalactic objects with an observed luminosity outshining their host galaxy. 
Their non-thermal continuum emission extending from radio band to X rays or $\gamma$ rays suggest underlying relativistic emission mechanisms.
AGN have been subject to comprehensive studies since their discovery, leading to classification and unification schemes based on multiwavelength observations and polarization measurements~\cite{urry95}.
In this frame, AGN are classified as either radio-quiet or radio-loud, and blazars constitute a subclass of the latter, with their jet axis oriented close to the observer's line of sight.
This particular orientation combined with relativistic beaming gives rise to prominent observational features in blazars, such as anisotropic radiation, rapid variability, high polarization and superluminal motion.
Blazars are divided into two subclasses, flat spectrum radio quasars (FSRQ) and BL Lacertae objects (BL Lac). 
FSRQs are characterized by a broad line emission region, depicted with strong lines in their optical spectra, that are not present in BL Lacs.
\par
The broadband spectral energy distribution (SED) of blazars exhibit a two-component structure, with a low-energy component peaking around IR to UV band and a high-energy one around X-ray to $\gamma$-ray energies.
The underlying emission mechanism reponsible for low-energy component is believed to be synchrotron emission from relativistic electrons in the blazar jets.
On the other hand, the high-energy component could either have leptonic~\cite{maraschi92,dermer93} or hadronic~\cite{aharonian00,mucke03} origin. 
Instrumental selection effects brought up two distinct classes for BL Lacs as radio-selected (RBL) or X-ray-selected (XBL). 
This instrumental classification was later replaced with a more physical one based on radio to optical and optical to X-ray spectral indices, introducing the terminology of \emph{high-} and \emph{low-frequency-cutoff} BL Lac objects (HBLs and LBLs)~\cite{padovani95}. 
Another approach based on the peak frequency of the synchrotron component of the SED is also used to define the same classification (\emph{high-} and \emph{low-frequency-peaked} BL Lacs as HBLs and LBLs). 
Detection of intermediate objects (IBLs) between these two observationally distinct groups has made it more plausible that BL Lac objects constitute a continuum rather than a discrete sequence~(e.g.,\cite{nieppola08}). 
\par
Mkn 421 was the first blazar and extragalactic object detected as a very high-energy (VHE; $E>100$\,GeV) $\gamma$-ray emitter with the Whipple telescope in 1992~\cite{punch92}.
Since then, different candidate selection methods have been applied to radio, X-ray or high-energy (HE; $E>100$\,MeV) $\gamma$-ray blazar data in the aim of finding new TeV blazars~\cite{costamante02,perez03,behera09}.
To date, 40 blazars have been detected in the TeV sky\footnote[1]{http://tevcat.uchicago.edu/}, with a census consisting of 29 HBLs, 4 IBLs, 4 LBLs and 3 FSRQs.

\section{SOURCE SAMPLE}
Our blazar sample contains all blazars with a published VHE spectrum before February 2011, with a total of 26 sources (see Table~\ref{sample}). 
This includes 19 HBLs, 3 IBLs, 2 LBLs and 2 FSRQs. 
Seven of these blazars were detected with EGRET and 23 of them are in the \emph{Fermi} 2-year catalog. 
More than half of the sample have been detected multiple times in the VHE band. 
These multiple detections extending over several years and obtained mostly with different instruments suggest that spectral variability in the VHE band is a common property for VHE blazars. 
Even though no general pattern has been established for VHE variability, several sources have been observed to have a flux increase up to a few times their baseline emission~\cite{ver_wCom_flare,magic_501_flare,2155_flare}, occasionally accompanied by a change in spectral index~\cite{magic_501_flare} and minute-scale flux doubling times~\cite{magic_501_flare,2155_flare}.
\begin{table}[h]
\centering
\begin{small}
\begin{tabular}{|c|ccc|}
\hline
Name 		& SED type 	& z 	 & \emph{Fermi} state\\
\hline
\hline
RGB J0152+017	& HBL 		& 0.080  & average	\\
\hline
3C 66A 		& IBL		& 0.444? & simultaneous	\\
\hline
1ES 0229+200 	& HBL 		& 0.140	 & average	\\
\hline
1ES 0347-121 	& HBL 		& 0.188  & average	\\
\hline
PKS 0548-322	& HBL		& 0.069	 & average	\\
\hline
RGB J0710+591	& HBL		& 0.125	 & simultaneous	\\
\hline
S5 0716+714	& LBL		& 0.300	 & high		\\
\hline
1ES 0806+524	& HBL		& 0.138	 & average	\\
\hline
1ES 1011+496	& HBL		& 0.212	 & high		\\
\hline
1ES 1101-232	& HBL		& 0.186	 & average	\\
\hline
Markarian 421	& HBL		& 0.031	 & average	\\
\hline
Markarian 180	& HBL		& 0.046	 & average	\\
\hline
1ES 1218+304	& HBL		& 0.182	 & simultaneous	\\
\hline
W Comae		& IBL		& 0.102  & high		\\
\hline
4C +21.35	& FSRQ		& 0.432	 & simultaneous	\\
\hline
3C 279		& FSRQ		& 0.536	 & high		\\	
\hline
PKS 1424+240	& IBL		& 0.260? & high		\\
\hline
H 1426+428	& HBL		& 0.129	 & average	\\
\hline
PG 1553+113	& HBL		& 0.4?	 & high		\\
\hline
Markarian 501	& HBL		& 0.034	 & low		\\
\hline
1ES 1959+650	& HBL		& 0.048	 & low		\\
\hline
PKS 2005-489	& HBL		& 0.071	 & average	\\
\hline
PKS 2155-304	& HBL		& 0.117	 & simultaneous	\\
\hline
BL Lacertae	& LBL		& 0.069	 & high		\\
\hline
1ES 2344+514	& HBL		& 0.044	 & average	\\
\hline
H 2356-309	& HBL		& 0.167	 & average	\\
\hline
\end{tabular}
\end{small}
\caption{{\small VHE blazar sample. No secure redshift is available for 3C 66A, PKS 1424+240 and PG 1553+113. The \emph{Fermi} states "low" and "high" are as described in the text. In some cases simultaneous data were available, and in the remaining cases 27-month time-averaged \emph{Fermi} data were used.}}
\label{sample}
\end{table}
\par
The first 27 month of \emph{Fermi} data and archival VHE spectra published before February 2011 were used to construct combined GeV-TeV SEDs. 
These include six data sets where VHE data overlaps with the \emph{Fermi} era (RGB J0710+591, 1ES 1218+304, PKS 1424+240, PKS 2155-304 and two different measurements for 3C 66A). 
The remainder of the VHE data were taken before the \emph{Fermi} mission. 
All VHE spectra were corrected for the extragalactic background light absorption using~\cite{dominguez2011}.

\section{\emph{Fermi} ANALYSIS}
The fact that most of the GeV and TeV data are not contemporaneous introduces caveats for the interpretation of the combined spectra. 
Moreover, \emph{Fermi} data represents an average state over a fairly long period, whereas the VHE spectra consist of day-scale ``snapshots", mostly taken during flares. 
As a solution to this problem, for bright-enough sources, the \emph{Fermi} data were split into ``low" and ``high" states as described below. 
Thus, non-contemporaneous GeV and TeV measurements were matched in a more realistic way than directly using all the time-averaged \emph{Fermi} data.
\par
\emph{Diffuse class} events with energy between 300\,MeV--100\,GeV from the first 27-month of \emph{Fermi} data (from 4 August 2008 to 4 November 2010) were used for the analysis.
For blazars that were observed during the \emph{Fermi} era, only the time periods of a few months that cover the corresponding VHE observations were selected. 
\par
\begin{figure}[th]
\centering
\includegraphics[width=70mm]{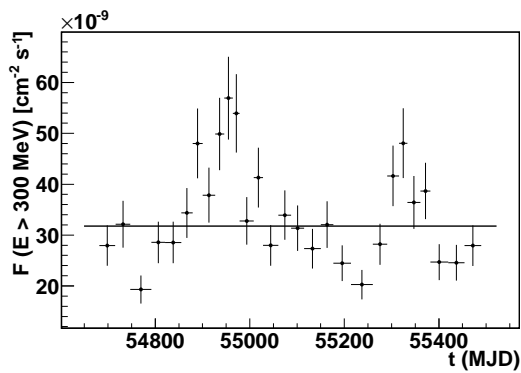}
\caption{Aperture photometry light curve of the HBL Mrk\,501. The solid line represents the average count rate, that is used as a reference point to split the data into low- and high-flux states.} 
\label{lc_mrk501}
\end{figure}

First, a 27-month light-curve analysis was performed for each source using an aperture photometry technique.
Events from a region of $1^{\circ}$ radius from the target location were selected and counts were plotted as a function of time, each time bin containing 49 counts. 
For sources with high statistics, low- and high- flux states were identified and separated using the average count rate as a threshold (see Figure~\ref{lc_mrk501} and Table~\ref{sample}).
\par
\begin{figure}[th]
\centering
\includegraphics[width=70mm]{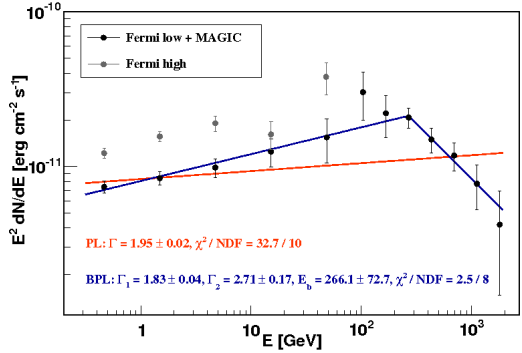}
\caption{Combined GeV-TeV SED of the HBL Mrk\,501 (black circles). The VHE spectrum belongs to a low flux state~\cite{anderhub2009a}. Therefore, the low-state \emph{Fermi} spectrum was selected for the combined GeV-TeV SED. A broken power law yields a better fit than a simple power law.} 
\label{sed_mrk501}
\end{figure}

Next, a spectral analysis was done for each data set. 
Events from a region of interest of $8^{\circ}$ were selected and analyzed with Fermi Science Tools v9r18p6\footnote{http://fermi.gsfc.nasa.gov/ssc/data/analysis/scitools/over view.html}, using instrument response functions P3\_V6\_DIFFUSE. 
Sources from the 1FGL catalog~\cite{abdo2010}, bright spots with test statistics $>25$ and standard galactic and isotropic diffuse emission background components\footnote{http://fermi.gsfc.nasa.gov/ssc/data/access/lat/Backgroun dModels.html} within the region of interest were included in the source model files. 
Spectral points were calculated using an unbinned maximum-likelihood analysis technique~\cite{cash79,mattox96}. 
Figure~\ref{sed_mrk501} shows the combined SED for the HBL Mrk\,501.

\section{IC COMPONENT}

\begin{figure}[th]
\centering
\includegraphics[width=70mm]{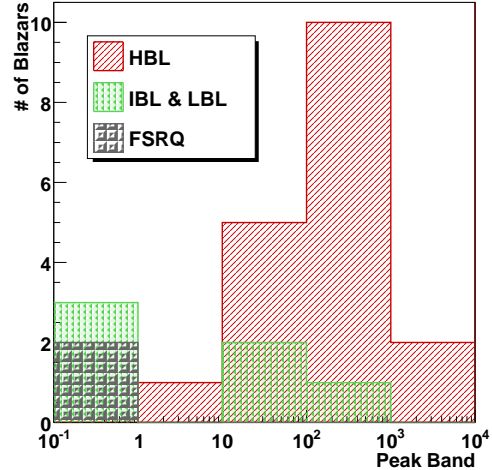}
\caption{Distribution of peak frequency bands.} 
\label{IC_peak_dist}
\end{figure}

The peak frequency of the inverse-Compton (IC) component is a salient parameter for describing blazar non-thermal continua and studying population trends. 
Systematic studies for measuring the IC peak frequency mostly suffer from the lack of statistics and simultaneous data~\cite{zhang2011}. In this work, we focus on finding the IC ``peak frequency band" rather than the ``peak frequency", using a model independent approach. For each blazar SED, we identify the energy decade in which the largest amount of power is emitted. 
Figure~\ref{IC_peak_dist} shows the distribution of peak frequency bands. 
We observe that the low-synchrotron-peaked objects (LSPs) have the maximum of their emission mostly below 1 GeV. 
On the other hand, high-synchrotron-peaked objects (HSPs) tend to peak in the TeV range. 

\section{INTERNAL ABSORPTION}
A power-law function often fails to describe the combined SEDs successfully. For most blazars, the fits are significantly improved with a broken power law (see Figure~\ref{sed_mrk501}) or an absorbed power law. 
Particularly, some LBL objects display absorption-like features in the 10-100 GeV range. 
To investigate this feature, we considered the following absorption scenarios from the broad line region (BLR)~\cite{poutanen2010} for the blazars S5~0716+714 (LBL), W Comae (IBL) and BL Lacertae (LBL):

\begin{itemize}
\item H I line (13.6 eV)
\item He II line (54.4 eV)
\item H I \& He II combined
\item Full BLR spectrum
\end{itemize}

\begin{figure}[th]
\centering
\includegraphics[width=70mm]{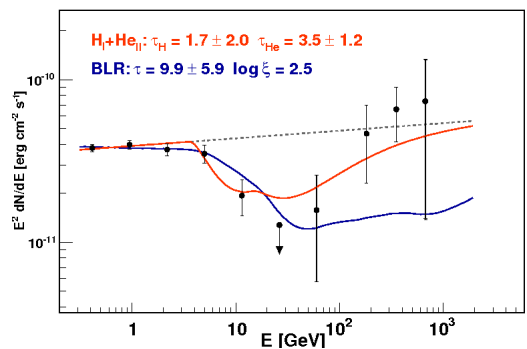}
\caption{\emph{Fermi} high-state + MAGIC~\cite{anderhub2009b} spectrum for the LBL S5 0617+714. 
The red curve represents H I + He II absorption only and the blue one the full BLR absorption. 
The power-law fit is shown with the dashed line.} 
\label{s5_0716_sed}
\end{figure}

\begin{figure}[th]
\centering
\includegraphics[width=70mm]{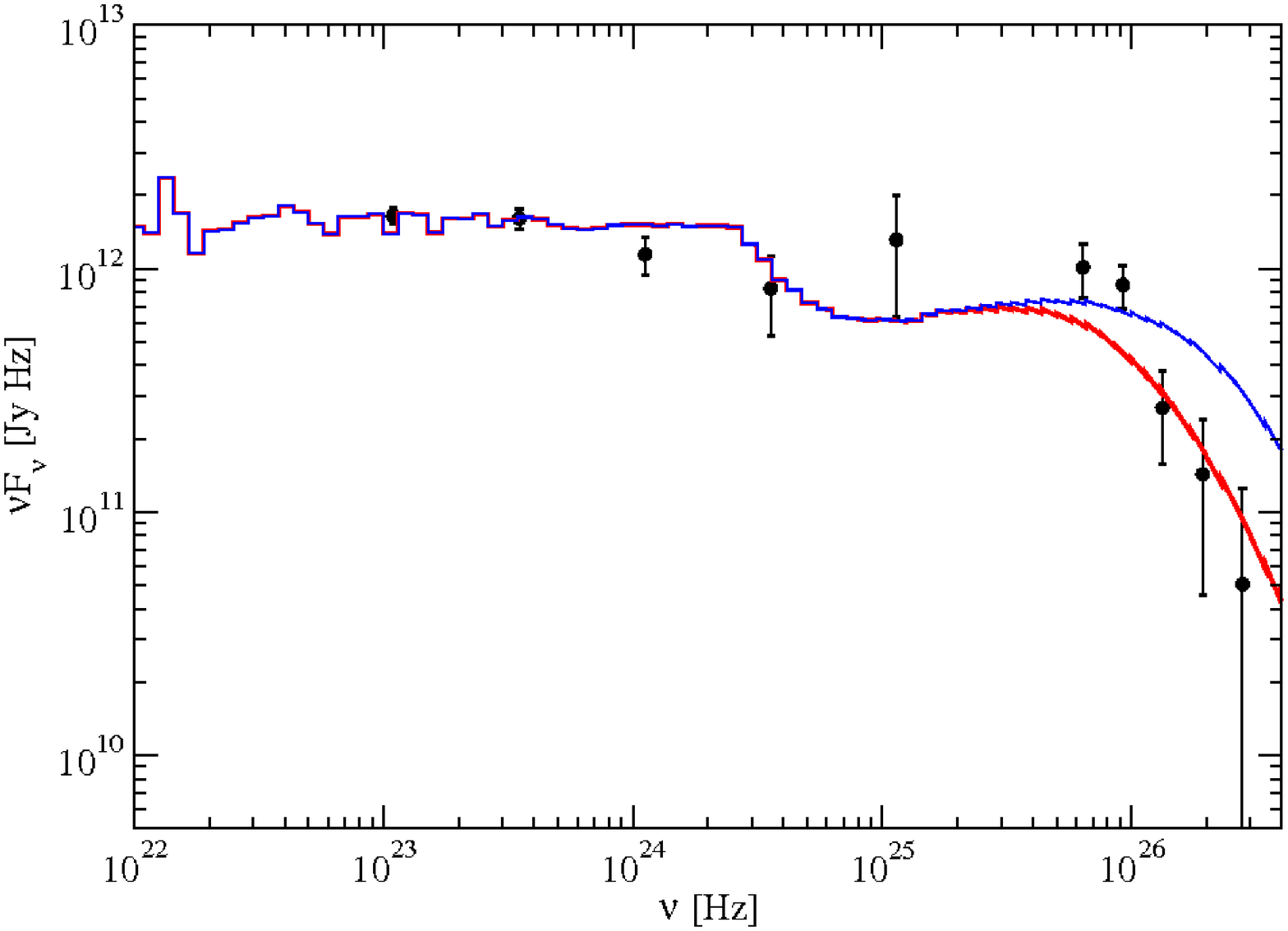}
\caption{\emph{Fermi} high-state + VERITAS~\cite{acciari2008} spectrum for the
LBL W Comae, with absorption of a pure power law (blue) and a power law with exponential cut-off (red), by the full BLR emission, with cascading effects taken into account.} 
\label{wComae_sed}
\end{figure}

In all the cases, absorption from the He II complex seems to be dominant. 
A double-absorption scenario does not provide an improved fit over the He II single-line absorption.
On the other hand, the full BLR absorption underproduces the TeV emission (see Figure~\ref{s5_0716_sed}).
We also modeled a full BLR absorption combined with cascade emission from the $\gamma$-$\gamma$ absorption~\cite{roustazadeh2010}. 
No significant contribution from cascading effects in the Fermi band is expected (see Figure~\ref{wComae_sed}).

\section{SUMMARY} 

In this paper we give some results from our ongoing GeV-TeV blazar population studies.
We analyzed 27 months of \emph{Fermi} data and combined it with archival TeV data. 
We see absorption-like features mostly in LBL spectra, these features are best described as BLR absorption from the He II complex. 
In addition, we observe that the peak frequency in the GeV-TeV region follows the synchrotron peak frequency trend.
Complete results will be discussed in a forthcoming paper.

\bigskip 
\begin{acknowledgments}
This work was supported in part by the NSF grant Phy-0855627 and NASA grant NNX09AU14G. The authors thank Juri Poutanen for providing a set of
BLR absorption templates.
\end{acknowledgments}

\bigskip 

\end{document}